

\magnification=\magstep1

\hsize = 33pc
\vsize = 45pc
\font\grand=cmbx10 at 14.4truept

\def\bs{\baselineskip=14.5pt}

\pageno=0
\def\folio{
\ifnum\pageno<1 \footline{\hfil} \else\number\pageno \fi}

\baselineskip=12pt
\phantom{DIAS}
\rightline{ BONN--HE--93--25 \break}
\rightline{ DIAS--STP--93--13 \break}
\rightline{ July 1993\break }
\rightline{ hep-th/9307190\break}
\vskip 1.0truecm

\bs
\tolerance=8000
\parskip=5pt

\def\pa{\partial}
\def\de{\delta}

\def\G{{\cal G}}
\def\K{{\cal K}}

\def\W{{\cal W}}

\def\O{{\cal O}}
\def\L{{\cal L}}
\def\S{{\cal S}}

\def\la{\langle}
\def\ra{\rangle}

\centerline
{\grand The vacuum preserving Lie algebra  of a classical
 ${\cal W}$-algebra}

\vskip 0.8truecm
\centerline{L. Feh\'er\footnote*{
An Alexander von Humboldt Fellow. On leave from
Bolyai Institute of Szeged University, H-6720 Szeged, Hungary.}
}
\medskip
\centerline{\it Physikalisches Institut der Universit\"at Bonn}
\centerline{\it Nussallee 12, 53115 Bonn, Germany}
\vskip 0.6truecm
\centerline{L. O'Raifeartaigh and  I. Tsutsui}
\medskip
\centerline{\it Dublin Institute for Advanced Studies}
\centerline{\it 10 Burlington Road, Dublin 4, Ireland}

\vskip 1.6truecm
\centerline{\bf Abstract}
\vskip 0.2truecm

We simplify and generalize
an argument due to Bowcock and Watts
showing that one can associate a finite Lie algebra
(the `classical vacuum preserving algebra')
containing the M\"obius $sl(2)$ subalgebra to any
classical $\W$-algebra.
Our construction is based on a kinematical analysis of
the Poisson brackets of quasi-primary fields.
In the case of the $\W_\S^\G$-algebra
constructed through the Drinfeld-Sokolov reduction
based on an arbitrary $sl(2)$ subalgebra $\S$
of a simple Lie algebra $\G$,
we exhibit a natural  isomorphism between
this finite Lie algebra and  $\G$ whereby the M\"obius
$sl(2)$ is identified with $\S$.

\vfill\eject

\centerline{\bf 0. Introduction}
\medskip

The classification of two dimensional conformal field
theories (CFT) is an outstanding problem in theoretical
physics, which is important  for  string theory,
statistical physics, integrable systems, and various
branches of mathematics.
A subproblem to this is the problem of classifying the
$\W$-algebras, that is the possible polynomial
extensions of the chiral conformal algebra, the Virasoro
algebra, by chiral primary fields.
There is by now a large literature on this problem
(see e.g.~[1] and references therein), whose study was
initiated by A. B. Zamolodchikov [2] by constructing unexpected examples
of extended conformal algebras, in particular the $W_3$-algebra
existing for continuously variable Virasoro centre.
The classical  version of the $W_3$-algebra
was  subsequently recognized [3] as
the second Poisson bracket (PB) algebra of the simplest
generalized KdV hierarchy (the Boussinesq hierarchy),
generalizing the relationship between the Virasoro algebra and
the KdV hierarchy.
Then it became clear that the second Poisson
structures of the generalized KdV  hierarchies constructed
by means of a Hamiltonian reduction method
by Drinfeld and Sokolov [4] all qualify as `classical
$\W$-algebras'.
This led to extensive studies [5,6,7]
on the quantization of the
Hamiltonian reductions of the affine Kac-Moody
PB algebras considered by Drinfeld and Sokolov,
and for a hunt after new  $\W$-algebras and
integrable systems obtained by
generalizations of the Drinfeld-Sokolov construction.

Recently there has been remarkable progress in this field.
First a classical $\W$-algebra was associated to each embedding
of the Lie algebra $sl(2)$ into a simple Lie algebra [8]
(see also [9]).
This was done by a  natural generalization of the Kac-Moody reduction
considered by Drinfeld and Sokolov,
using its interpretation in terms of the principal $sl(2)$ embedding
found in [10].
More recently,  an elegant quantization of this set of $\W$-algebras
(called $\W_\S^\G$-algebras where $\S\subset \G$ is the $sl(2)$ embedding)
was found [11,12] in the quantum Hamiltonian reduction (BRST) framework.
At the same time, a number of arguments
was put forward [13]
indicating that there are probably no
other `nice'  $\W$-algebras that one could obtain in the
Hamiltonian reduction framework since the  $\W_\S^\G$-algebras
appear to  exhaust the $\W$-algebras
resulting from Drinfeld-Sokolov type reduction.
For example, it was shown that the spectrum of conformal
weight  $\Delta\geq {3\over 2}$ generators in any such $\W$-algebra
must always be the same as in a corresponding
$\W_\S^\G$-algebra.

The above results on the construction and classification
of $\W$-algebras were obtained in the Hamiltonian  reduction approach,
but there exists also a more abstract argument pointing
to the crucial r\^ole of $sl(2)$ embeddings in this context.
This is due to Bowcock and Watts [14], who studied the classification
of `reductive' quantum $\W$-algebras.
These are operator product algebras (meromorphic
conformal field theories
[15]) generated by a Virasoro field and a finite set of
chiral
primary fields with half-integral conformal weights $\Delta \geq 1$ that
exist for a continuous range of the Virasoro centre $C$
and admit a classical ($C\rightarrow \infty$)
limit (see [14] for precise definition).
The corresponding classical $\W$-algebra is then a PB algebra
of fields $W^a(z)$ ($a=1,\ldots, N$) on the circle
$S^1$ ($\vert z\vert =1$)  of the type
$$
\{ W^a(z), W^b(w)\} =\sum_{k\geq 0}P^{ab}_k
\left(W^1(w),\ldots, W^N(w)\right) \pa^k_z \delta(z-w)\,,
\eqno(0.1)$$
where the sum is finite,
the $P^{ab}_k(W)$'s are differential polynomials or constants,
and one of the fields is a Virasoro density with respect to which
the others are primary fields.
The idea of Bowcock and Watts [14] is to consider
the space $\tilde \G$ of
`vacuum preserving Fourier modes'
given by
$$
\tilde \G:={\rm span}\{\,W^a_m\vert\, \vert m\vert \leq
 (\Delta_a-1)\,\}\,,
\eqno(0.2)$$
and define a Lie algebra structure on $\tilde \G$  by truncating
(0.1) to its linear part (for the Fourier mode convention, see (1.7)).
Then they  showed that this procedure
yields a finite Lie algebra by using the commutator formula
[16,17] of chiral quasi-primary fields in meromorphic CFT.
The importance of the construction derives from
the fact that $\tilde \G$ automatically contains the
M\"obius $sl(2)$ subalgebra $\tilde \S$
spanned by the vacuum preserving modes of the Virasoro.
This immediately implies that
the conformal spectrum of the $\W$-algebra
is encoded in the spin content of the decomposition of $\tilde \G$
into $\tilde \S$ irreps (which are simply the vacuum preserving
modes of the fields).
This means that the possible conformal spectra are determined
by the possible $sl(2)$ embeddings into finite Lie algebras,
and since it  was also argued [14] that the finite Lie algebra
$\tilde \G$ is necessarily a reductive (semisimple + abelian)
Lie algebra the $sl(2)$ embeddings are given by
Dynkin's list [18].

The main purpose of the present note is to point out
that the above argument actually applies to the classification
of classical $\W$-algebras directly, without
viewing them as the classical limit of a quantum
$\W$-algebra as was done in [14].
By using only some simple kinematical properties of Poisson brackets,
we shall show that one can  associate the finite Lie algebra
$\tilde \G$ (the `classical vacuum preserving algebra')
to any classical $\W$-algebra.
This means for example that
the result about the possible conformal spectra
applies directly to the classification of classical $\W$-algebras,
and then `a fortiori' to the quantum $\W$-algebras having a
classical limit.
Our purely classical derivation is not only a
simplification, but in this way we also obtain
a certain generalization of the  Bowcock-Watts result
since we do not use various,
not always satisfied, properties of an underlying meromorphic CFT
required in [14]
(e.g.~the positivity and integral conformal weight assumptions).
After presenting this,
in Section 2 we shall consider the $\W_\S^\G$-algebras and
show that the `vacuum preserving data' $(\tilde S , \tilde \G)$
can in this case be naturally identified
with the `Drinfeld-Sokolov data' $(\S, \G)$.
This result was derived in [14] in the integral weight case.
The derivation we present is different,
and we shall also cover the  half-integral case.

\bigskip

\centerline{\bf 1. Finite Lie algebra from PB algebra of quasi-primary
fields}
\medskip

Suppose that we have some finite set of
quasi-primary fields $W^a(z)$
defined on the circle $S^1$, with half-integral weights
$\Delta_a \geq 1$ with respect to an action of the
M\"obius group implemented through Poisson bracket:
$$\eqalign{
\{ L_m, W^a(z)\}&=z^m(z\pa_z + (m+1)\Delta_a ) W^a(z),\cr
\{ L_m, L_n\}& =(m-n) L_{m+n}\,,
\qquad m,n =  0,\pm 1 .\cr}
\eqno(1.1)$$
Suppose that the PB's of the fields $W^a(z)$ close in
a local, differential polynomial way.
More precisely, suppose that we have
$$
\{ W^a(z),W^b(w)\} =g_{ab}\de^{(\Delta_a +\Delta_b -1)}
+\sum_c \sum_{n=0}^{\Delta(abc)-1} A^{ab}_c(n) (\pa^n W^c)(w)
\de^{(\Delta(abc)-1-n)}+\cdots\,,
\eqno(1.2)$$
where the dots stand for terms  quadratic and higher order
in the $W$'s; the summation is on those $c$ for which
$\Delta(abc):=\Delta_a +\Delta_b-\Delta_c \in {\bf N}$;
$\de^{(m)}:=\pa_z^m \de(z-w)$;
and if $(\Delta_a +\Delta_b)\notin {\bf N}$ then $g_{ab}=0$.
We are interested in the  coefficients
$g_{ab}$ and $A^{ab}_c(n)$.
By taking the PB of eq.~(1.2) with the M\"obius generator
$L_1$ using (1.1),
and comparing the two sides of the resulting equation
up to linear terms in the $W$'s,
one obtains that $g_{ab}$ can be nonzero
only for fields of equal weights,
$$
g_{ab}=g_{ab}\delta_{\Delta_a,\Delta_b}\,,
\eqno(1.3)$$
and  also obtains a recursion relation for the $A^{ab}_c(n)$.
This leads to the following result:
$$
A^{ab}_c(n)={(1+\Delta_c-\Delta_a-\Delta_b)_{(n)}
(\Delta_a -\Delta_b+\Delta_c)_{(n)}\over n! (2\Delta_c)_{(n)} }
A^{ab}_c(0)\,,
\eqno(1.4)$$
where for any  $x\in {\bf C}$ and $n\in {\bf N}$ we
define
$$
(x)_{(n)}:=\prod_{m=0}^{n-1} (x+m)\,,
\qquad\hbox{and}\qquad
(x)_{(0)}:=1\,.
\eqno(1.5)$$

The next step is to express (1.2) in Fourier
modes by using eqs.~(1.3-4) together with
$$
\oint dz f(z) \de(z-w) =f(w)\,,
\qquad
\delta(z-w)= {1\over 2\pi {\it i}} \sum_{m\in {\bf Z}}
z^m w^{-(m+1)}\,,
\eqno(1.6)$$ and the modes defined by
$$
W^a(z):=\sum_{m\in {\bf Z}+\Delta_a} W^a_m z^{-m-\Delta_a}\,.
\eqno(1.7)$$
One then finds the following formula:
$$\eqalign{
\{ W^a_m , W^b_n\} =&
{{(-1)^{2\Delta_a-1}}\over {2\pi i}}
\left( \prod_{k=-(\Delta_a-1)}^{(\Delta_a-1)} (m +k) \right)
g_{ab}\delta_{\Delta_a,\Delta_b}\delta_{m,-n}
\cr
&+\sum_{\{ c: \Delta(abc)\in {\bf N}\}}
B^{ab}_c P(m,n;\Delta_a,\Delta_b,\Delta_c) W^c_{m+n}\cdots\,,\cr}
\eqno(1.8)$$
with
$$
B^{ab}_c:= {{(-1)^{\Delta(abc)-1}}\over {2\pi i}} (\Delta(abc)-1)!\,
 A^{ab}_c(0)\,,
\qquad
\Delta(abc)=\Delta_a +\Delta_b-\Delta_c\geq 1\,,
\eqno(1.9)$$
and
$$\eqalign{
P&(m,n;\Delta_a,\Delta_b,\Delta_c)=\cr
&\sum_{r=0}^{\Delta(abc)-1} {(-1)^r\over r!}
{m+\Delta_a -1\choose \Delta_a +\Delta_b -\Delta_c -1 -r}
{(\Delta_a-\Delta_b+\Delta_c)_{(r)}
(m+n+\Delta_c)_{(r)}\over (2\Delta_c)_{(r)}}\,.\cr}
\eqno(1.10)$$

As an aside, we should at this point
note that the same polynomials $P$ in
(1.10) appear in the commutator
formula of quasi-primary fields in the quantum case [16,17],
simply because they express purely kinematical information.
To derive the commutator formula one could proceed  in the
quantum case similarly as above,
by demanding the covariance of the operator product
of quasi-primary fields under the M\"obius group.
This derivation appears simpler than the ones given
in refs.~[16,17], which inspired the above consideration.

We now consider the \lq linearized vacuum
preserving
PB algebra' (vpa for short\footnote*{Although there is
obviously no vacuum
in the classical case we keep the terminology
vpa to maintain
contact with previous work [14].})
defined  by
$$
[W^a_m, W_n^b]:=\sum_c
B_c^{ab} P(m,n;\Delta_a,\Delta_b,\Delta_c) W^c_{m+n}\,,
\qquad \vert m\vert \leq (\Delta_a -1)\,,\
\vert n\vert \leq (\Delta_b-1)\,,
\eqno(1.11)$$
{\it i.e.}, we drop the quadratic and higher order
terms on the r.h.s. of (1.8).
It is not hard to see that this formula provides a PB
on the `vp modes' $\{ W^a_m\}$, $\vert m\vert \leq (\Delta_a -1)$.
Indeed, if $W^a_m$ and $W^b_n$ are vp modes
but $W^c_{m+n}$ is not so, {\it i.e.}, $\vert m+n\vert>(\Delta_c-1)$,
then
$$
P(m,n;\Delta_a,\Delta_b,\Delta_c)=0\,.
\eqno(1.12)$$
This follows from the fact that
$$
P(m,n;\Delta_a,\Delta_b,\Delta_c)\sim C^{j_a j_b j_c}_{mn}
\quad\hbox{for}\quad
\vert m\vert \leq j_a,\, \vert n\vert \leq j_b\,,
\eqno(1.13)$$
where $j_a:=\Delta_a-1$ (etc.), and $C^{j_a j_b j_c}_{mn}$ is the
standard Clebsch-Gordan coefficient of $su(2)$.
This was verified in [17] from the explicit  formula (1.10).
A noncomputational argument yielding (1.12-13) would be to
combine the fact that the vp modes of a
quasi-primary field carry a
finite dimensional irreducible representation
(irrep) of the M\"obius $sl(2)$ with the
fact that the set of such irreps is stable under tensor products,
and the PB of the vp modes behaves like a tensor product
when acted on by the M\"obius generators.
(We sketched the derivation of (1.10)
above since it contains more information.)
After establishing that only vp modes appear on the r.h.s. of (1.11),
one easily sees that the Jacobi identity is satisfied for the linearized
PB (1.11) as a consequence of the Jacobi identity of the
original PB (1.8) on account of the absence of
central term between a vp mode and an arbitrary other mode in (1.8).
In conclusion,  one can associate the finite dimensional
Lie algebra $(\tilde \G, [\ ,\ ])$ given by (0.2), (1.11)
to any  PB algebra of fields of the type given by (1.1-2).

In particular, if one of the fields in (1.2)
 is the Virasoro density whose vp modes are the
 M\"obius modes $\{ L_{-1}, L_0, L_1\}$,
then the vpa $\tilde \G$ automatically comes equipped
with the preferred $sl(2)$ subalgebra
$\tilde \S:={\rm span}\{\,L_{-1}, L_0, L_{+1}\,\}$.
In this case, by construction $\tilde \G$
decomposes  under $\tilde \S$ into the direct sum
of the spin $j_a:=(\Delta_a-1)$ irreps
spanned by the vp modes of the $W^a$.
This means that
the conformal spectrum $\Delta_a$
of the generating fields $W^a(z)$ of a classical $\W$-algebra
is always determined
by an $sl(2)$ decomposition of a finite Lie algebra and,
a fortiori, the same can be said about those quantum
$\W$-algebras that admit a classical limit [14].
Thus the question of classifying the possible conformal
spectra of such $\W$-algebras
reduces to the classification of $sl(2)$ embeddings,
which is known if
the vpa is a reductive (semisimple+abelian) Lie algebra [18].
Moreover,   $\W$-algebras
with any such `in principle possible' conformal spectrum
can actually be easily constructed by Drinfeld-Sokolov reduction.

\bigskip

\centerline{\bf 2. The vpa of $\W_\S^\G$ is $\G$}
\medskip

Let $\G$ be a finite dimensional simple Lie algebra and
$\S={\rm span}\{ M_-, M_0, M_+\}\subset \G$ an $sl(2)$ subalgebra,
$$
[M_0, M_\pm]=\pm M_\pm\,,
\qquad
[M_+, M_-]=2M_0\,.
\eqno(2.1)$$
To this data one can associate a natural $\W$-algebra, designated
as $\W_\S^\G$, and construct its vpa along the lines
of the previous section.
This gives a finite Lie algebra $\tilde \G$, endowed
with the M\"obius embedding $\tilde \S \subset \tilde \G$.
The $sl(2)$ spins in
the decomposition of $\G$ under $\S$ coincide with those
in the decomposition of $\tilde \G$ under $\tilde \S$ since
both sets of $sl(2)$ spins is related
to the set  of conformal weights of $\W_\S^\G$
by a shift by $1$ (as is obvious for $\tilde S\subset \tilde \G$ and
 well known for $\S\subset \G$).
Thus one expects $(\S,\G)$ to be isomporphic to
$(\tilde \S,\tilde \G)$.
Indeed, this has been shown by Bowcock and Watts [14]
in the case of an integral $sl(2)$ embedding, by a rather
involved computation.
The purpose of this section is to give a simple proof
covering half-integral $sl(2)$ embeddings as well.
We first recall the definition of the $\W_\S^\G$-algebra [8,9].

The $\W_\S^\G$-algebra is a reduction of the `KM Poisson bracket
algebra' constructed as follows.
We first consider the space $\K$ of $\G$-valued smooth functions
(`currents')
$J(z)$ defined on the circle $S^1$.
We let the space $\K$ carry the Poisson bracket
$$
\{
\langle \alpha , J(z)\rangle \,,\,\langle \beta , J(w)\rangle \}
=\langle [\alpha,\beta],J(z)\rangle\delta (z-w)
+  \langle \alpha,\beta \rangle \delta^{\prime}(z-w)\,,
\quad  \alpha,\beta \in \G\,,
\eqno(2.2)
$$
where $\la\ ,\ \ra$ is the invariant scalar product on $\G$.
For our purpose it is  useful to introduce the
`shifted current' $j(z):=J(z)-M_-$ in terms of which (2.2) reads
$$
\{
\langle \alpha , j(z)\rangle \,,\,\langle \beta , j(w)\rangle \}
=\left(\langle [\alpha,\beta],j(z)\rangle +\omega(\alpha,\beta)\right)
\delta (z-w)
+ \langle \alpha,\beta \rangle \delta^{\prime}(z-w)\,,
\eqno(2.3)
$$
where $\omega$ is the 2-form  on $\G$  given by
$\omega(\alpha,\beta):=\la M_-,[\alpha,\beta]\ra$.
We then impose the constraints
$$
\phi_{\tau_i}(z):=\la \tau_i,j(z)\ra =0\,,
\eqno(2.4)$$
where $\{\tau_i\}$ is a basis of the subspace
$[M_+,\G]\subset \G$.
The corresponding constraint surface
is the affine subspace $\K_{\rm hw}\subset \K$
consisting of currents of the  following
($sl(2)$ highest weight) form
$$
J(z)=M_-+j_{\rm hw}(z)\,,
\qquad
j_{\rm hw}(z)\in {\rm Ker} \left({\rm ad}_{M_+}\right).
\eqno(2.5)$$
The constraints (2.4) are second class and the
$\W_\S^\G$-algebra is just the Dirac bracket
algebra carried by the components of the constrained current
$j_{\rm hw}(z)$.
It is convenient to choose a basis
$\{ \sigma^{a,m}\}\subset \G$ adapted to the decomposition
of $\G$ under $\S$.
The index $a$ runs over the $sl(2)$ irreps and
for fixed $a$ we have $-l_a \leq m \leq l_a$, where $l_a$ is the spin
of the irrep, with the  normalization fixed according to
$$
\sigma^{a,m}=
\left({\rm ad}_{M_-}\right)^{l_a-m} (\sigma^{a,l_a})\,,
\qquad
[M_0, \sigma^{a,m}] = m \sigma^{a,m}\,.
\eqno(2.6)$$
Defining $W^a(z):=\la \sigma^{a,-l_a}, j_{\rm hw}(z)\ra$,
the Dirac bracket of our interest  is
$$\eqalign{
\bigl\{&  W^a(z),  W^b(w) \bigr\}^*=\Bigl(
\bigl\{ \la \sigma^{a,-l_a} , j(z)\ra, \la \sigma^{b,-l_b}, j(w)\ra \bigr\}\cr
&-\sum_{i,k} \oint \oint du\,dv
\bigl\{ \la \sigma^{a,-l_a}, j(z)\ra, \phi_{\tau_i} (u)\bigr\}
D^{-1}_{ik}(u,v;j)
\bigl\{ \phi_{\tau_k} (v),\la \sigma^{b,-l_b}, j(w)\ra \bigr\}
\Bigr)\vert_{\K_{\rm hw}}, \cr}
\eqno(2.7)$$
where $D^{-1}_{ik}(u,v;j)$ is the inverse of
$D_{ik}(u,v;j):=\{ \phi_{\tau_i}(u),\phi_{\tau_k}(v)\}$, {\it i.e.},
$$
\sum_k \oint du D_{ik}(z,u;j) D^{-1}_{km}(u,w;j)=\delta_{im}\,
 \delta(z-w)\,.
\eqno(2.8)$$
Eq.~(2.7) defines a differential polynomial algebra since
$D^{-1}_{ik}(u,v;j)$ is a linear combination of
$\delta(u-v)$ and its derivatives
with coefficients that are differential polynomials in $j(u)$.
The field $W^a(z)$ is quasi-primary with weight $\Delta_a=l_a+1$
with respect to the Virasoro density
$$
L(z)=\la M_-, j_{\rm hw}(z)\ra + {1\over 2} \la j_{\rm hw}(z),
 j_{\rm hw}(z)\ra,
\eqno(2.9)$$
whose two terms are commuting Virosoro densities (under the
Dirac bracket).
Notice also that the Virasoro (2.9) is the restriction of the
Sugawara formula to the constraint surface (2.5).

Having recalled the definition of $\W_\S^\G$,
we are now in a position
to study its vpa $(\tilde \G, [\ ,\ ])$, which is
obtained (cf.~(1.8) and (1.11)) by
linearizing (2.7) as
$$
\{ W_m^a , W_n^b\}^* := [ W_m^a, W_n^b] + \O(j_{\rm hw}^2)\,,
\eqno(2.10)$$
for the vp modes
$$
W_{m}^a  :={1\over {2\pi i}} \oint dz\, W^a(z) z^{m+l_a}\,,
\qquad  -l_a \leq m \leq l_a\,,
\eqno(2.11)$$
and similarly for $b$.
In (2.10)  $\O(j_{\rm hw}^2)$  denotes
the quadratic and higher order terms in the Dirac ($\W$-algebra) bracket
of the vp modes.
We shall prove that the following correspondence,
$$
\sigma^{a,m}\Longleftrightarrow {\widehat W}^a_m\,,
\quad\hbox{with}\quad
{\widehat W}^a_m:={{2\pi i}\over {(m+l_a)!}} W^a_m\,,
\eqno(2.12{\rm a})$$
is an isomorphism between the original Lie algebra  $\G$
and the vpa $\tilde \G$,
$$
[\sigma^{a,m}, \sigma^{b,n}]
\Longleftrightarrow
[{\widehat W}_m^a, {\widehat W}_n^b]\,.
\eqno(2.12{\rm b})$$

We start by noting that, for any test function $f(z)$,
we have the $\W$-charge
$$
W_f^a [j_{\rm hw}]:=\oint dz \la \sigma^{a,-l_a}  , j_{\rm hw}(z)\ra f(z)\,,
\eqno(2.13)$$
which is a function
on $\K_{\rm hw}$ (the vp modes (2.11) being special cases).
We then define the corresponding  \lq improved $\W$-charge'
$$
\eqalign{
{\widetilde W}^a_f[j] &:= \oint dz \la \sigma^{a,-l_a}  , j(z)\ra f(z)
\cr
&\quad - \sum_{i,k} \oint\oint\oint \,du\,dv\,dz\;
 f(z) \bigl\{ \la \sigma^{a,-l_a}, j(z) \ra , \,
     \phi_{\tau_i}(u)\bigr\} \,D^{-1}_{i,k}(u,v;j_{\rm hw}) \,
 \phi_{\tau_k}(v)\,,
}
\eqno(2.14) $$
which is a function on
$\K$ that reduces to $W^a_f$ on $\K_{\rm hw}$
and (weakly) commutes with the constraints,
$$
\{ {\widetilde W}^a_f, \phi_{\rm \tau_i}(z)\}\vert_{\K_{\rm hw}}=0.
\eqno(2.15)$$
(For clarity, we note that $j_{\rm hw}$ in the argument
of $D^{-1}$ in (2.14)
denotes the projection of $j$ on
${\rm Ker}\left({\rm ad}_{M_+}\right)$,
given by the decomposition
$\G=[M_-,\G] +{\rm Ker}\left({\rm ad}_{M_+}\right)$).
 Since ${\widetilde W}^a_f[j]$ is a differential polynomial
in $j$, we can separate its  linear part,
denoted by $\L_f^a[j]$, as
$$
{\widetilde W}^a_f[j] =\L_f^a[j] +\O(j^2)\,.
\eqno(2.16)$$
We shall see below that the information contained in the linearized
Dirac bracket giving the vpa by (2.10) can be extracted
from the linear terms of the improved $\W$-charges.
Taking the linear part of (2.14) yields
$$
\eqalign{
{\L}&^a_f[j]:= \oint dz \la \sigma^{a,-l_a}  , j(z)\ra f(z)
\cr
&\quad - \sum_{i,k} \oint\oint\oint \,du\,dv\,dz\;
 f(z) \bigl\{ \la \sigma^{a,-l_a}, j(z) \ra  ,
     \phi_{\tau_i}(u)\bigr\}_{j=0}\,D^{-1}_{i,k}(u,v;j_{\rm hw}=0)
\, \phi_{\tau_k}(v)\,,
}
\eqno(2.17) $$
from which  we find the following  explicit formula:
$$
\L^a_f[j] = \sum_{k=0}^{2l_a}
\oint dz\, \la \sigma^{a, k-l_a},j(z) \ra
\partial^k f(z)\,.
\eqno(2.18)$$
As an alternative to inspecting  the kernel
$D_{ik}^{-1}(u,v; j_{\rm hw}=0)$,
the reader can verify this formula by checking that it satisfies
$$
\{  \L^a_f, \phi_{\rm \tau_i}(z)\}_{j=0}=0\,,
\eqno(2.19)$$
which is a consequence of eqs.~(2.15-16).

Equation (2.18) is the key to our considerations
since we now show that
it implies the required isomporhism (2.12).
For this we first  note that,
because of the identity $W^a_f[j_{\rm hw}]={\widetilde W}^a_f[j_{\rm hw}]$
and (2.15), we have
$$
\{ W^a_{f_1}, W^b_{f_2}\}^*[j_{\rm hw}] =\{ {\widetilde W}^a_{f_1},
{\widetilde W}^b_{f_2}\}[j_{\rm hw}]\,,
\eqno(2.20)$$
for any two test functions $f_1$, $f_2$.
In other words,
we can use the original PB instead of the Dirac bracket
if we replace the $\W$-charges by their `improved' versions.
(A similar trick is always available, at least locally, when
dealing with a Dirac bracket.)
On the other hand,  we have
$$
\{ \L_f^a, \O(j^2)\}[j]=\O(j^2)\,,
\quad\hbox{for}\quad
f(z) \sim z^{m+l_a}
\quad\hbox{with}\quad
\vert m \vert \leq l_a,
\eqno(2.21)$$
as is readily verified from (2.18) using (2.3).
By combining (2.20) with (2.16) and (2.21),
we see that the linearized bracket $[W_m^a, W_n^b]$ defined in (2.10)
is the same as the original Kac-Moody PB  of the
linear terms of the corresponding improved $\W$-charges.
More precisely, we have
$$
[{\widehat W}^a_m, {\widehat W}^b_n]=\{ \L_m^a, \L_n^b\} [j_{\rm hw}]\,,
\eqno(2.22)$$
where we used the notation introduced in (2.12{\rm a}) and defined
$$
\L^a_m:=\L^a_f \quad
\hbox{with}\quad
f= {{z^{m+l_a}}\over{(m+l_a)!}}\,.
\eqno(2.23)$$
Observe that one can rewrite $\L^a_m$ given by (2.18) and (2.23) as
$$
\L^a_m[j]= \oint dz \la {\widehat \sigma}^{a,m}, j(z)\ra\,,
\eqno(2.24{\rm a})$$
where
$$
{\widehat \sigma}^{a,m}:=g \sigma^{a,m} g^{-1}
\quad\hbox{with}\quad
g:=e^{z M_-}\,.
\eqno(2.24{\rm b})$$
{}From this and (2.3) it follows that if in the $\S$-basis $\G$ is
 given by
$$
[\sigma^{a,m}, \sigma^{b,n}] =
\sum_{c,k} f^{(a,m), (b,n)}_{(c,k)}\, \sigma^{c,k}
\eqno(2.25)$$
then we have
$$
\{\L^a_m,\L^b_n\} [j] =\sum_{c,k} f^{(a,m), (b,n)}_{(c,k)}\, \L^c_k[j]\,.
\eqno(2.26)$$
Combining this with (2.22) and
using that $\L_k^c[j_{\rm hw}]={\widehat W}_k^c$, we conclude that
(2.12)  gives indeed  the claimed isomorphism  between the vpa of
$\W_\S^\G$ and $\G$:
$$
[{\widehat W}^a_m, {\widehat W}^b_n]=
\sum_{c,k} f^{(a,m), (b,n)}_{(c,k)}\, {\widehat W}_k^c\,.
\eqno(2.27)$$

Under the isomorphism (2.12) the $sl(2)$ subalgebra
$\S={\rm span}\{ M_-, M_0, M_+\}\subset \G$
corresponds to the $sl(2)$ spanned by the vp modes
of $\la M_-, j_{\rm hw}\ra$, the first term in (2.9).
At the linearized level this $sl(2)$ is the same as the
M\"obius $sl(2)$ spanned by the vp modes of the full
Virasoro density $L$ in (2.9), and thus we have recovered
the $sl(2)$ embedding data used in the construction
of the $\W_\S^\G$-algebra from its linearized vacuum preserving algebra.

We finish by noting that it is an interesting open question
whether a classical $\W$-algebra
is completely determined by its vpa and centre in general.

\vskip 0.7cm
\noindent
{\bf Acknowledgements.}
L.F. wishes to thank H. Arfaei, A. Honecker
and W. Nahm for discussions and for comments on the manuscript.
He is also grateful  to the Alexander von Humboldt-Stiftung
for support.

\vfill\eject

\hsize = 33pc
\vsize = 46pc

\centerline{\bf References}
\smallskip

\item{[1]}
P. Bouwknegt and  K. Schoutens,  Phys. Rep. {\bf 223} (1993) 183.
\item{[2]}
A. B. Zamolodchikov,  Theor. Math. Phys. {\bf 65} (1985) 1205.
\item{[3]}
K. Yamagishi,   Phys. Lett. {\bf 205B} (1988) 466;
\item{}
P. Mathieu, Phys. Lett. {\bf 208B} (1988) 101;
\item{}
I. Bakas,  Phys. Lett. {\bf 213B} (1988) 313.
\item{[4]}
V. G. Drinfeld and V. V.  Sokolov,  Jour. Sov. Math. {\bf 30} (1984)
1975.
\item{[5]}
V. A. Fateev and S. L.  Lukyanov,  Int. J. Mod. Phys. {\bf A3}
 (1988) 507;
Sov. Sci. Rev. {\bf A} Phys. {\bf 15} (1990) 1.
\item{[6]}
M. Bershadsky and H. Ooguri,
Commun. Math. Phys. {\bf 126} (1989) 49.
\item{[7]}
B. L. Feigin and E. Frenkel,
Phys. Lett. {\bf 246B} (1990)  75;
\item{}
E. Frenkel, V. G. Kac and M. Wakimoto,
Commun. Math. Phys. {\bf 147} (1992) 295.
\item{[8]}
F. A. Bais, T.  Tjin and  P. van Driel,  Nucl. Phys. {\bf B357}
(1991) 632.
\item{[9]}
L. Feh\'er, L. O'Raifeartaigh, P. Ruelle, I. Tsutsui and A. Wipf,
Phys. Rep. {\bf 222} (1992) 1.
\item{[10]}
J. Balog, L. Feh\'er, P. Forg\'acs, L. O'Raifeartaigh and A. Wipf,
Ann. Phys. (N.~Y.) {\bf 203} (1990) 76.
\item{[11]}
J. de Boer and T. Tjin,
preprint THU-92-32, IFTA-28-92, hep-th/9211109;
preprint THU-93/05, IFTA-02-93, hep-th/9302006.
\item{[12]}
A. Sevrin and W. Troost, preprint LBL-34125, UCB-PTH-93/19,
 KUL-TF-93/21,
hep-th/9306033.
\item{[13]}
L. Feh\'er, L. O'Raifeartaigh, P. Ruelle and I. Tsutsui,
preprint BONN-HE-93-14, DIAS-STP-93-02, hep-th/9304125.
\item{[14]}
P. Bowcock and G. M. T.  Watts,  Nucl. Phys. {\bf B379} (1992) 63.
\item{[15]}
P. Goddard, in: Infinite Dimensional Lie Algebras and
Lie Groups, ed. V. G. Kac, World Scientific, 1989.
\item{[16]}
W. Nahm, in: Proc. Trieste Conf. on Recent Developments in
Conformal Field Theory, Trieste, October 1989;
Conformal Quantum Field Theories in Two Dimensions,
World Scientific, to appear.
\item{[17]}
P. Bowcock, Nucl. Phys. {\bf B356} (1991) 367.
\item{[18]}
E. B. Dynkin,  Amer. Math. Soc. Transl. {\bf 6 [2]} (1957) 111.

\bye